\documentclass[aps, prl, superscriptaddress, reprint, twocolumn, amsmath, amssymb, a4, notitlepage]{revtex4-1}
\usepackage{graphicx}
\usepackage{amsmath,amssymb}
\usepackage{dcolumn}
\usepackage{mathrsfs}
\usepackage{physics}
\usepackage[utf8]{inputenc}
\usepackage{color}

\begin{document}
	\title{Quantum light-matter interaction and controlled phonon scattering in a photonic Fano cavity}
\author{Emil V. Denning}
\affiliation{Department of Photonics Engineering, DTU Fotonik, Technical University of Denmark, Building 343, 2800 Kongens Lyngby, Denmark}
\author{Jake Iles-Smith}
\affiliation{Department of Physics and Astronomy, University of Sheffield, Sheffield, S3 7RH, United Kingdom}
\author{Jesper Mork}
\email{jesm@fotonik.dtu.dk}
\affiliation{Department of Photonics Engineering, DTU Fotonik, Technical University of Denmark, Building 343, 2800 Kongens Lyngby, Denmark}

\begin{abstract}
The Fano effect arises from the interference between a continuum of propagating modes and a localised resonance. By using this resonance as one of the mirrors in an optical cavity, a localised mode with a highly asymmetric line shape is obtained. Placing a single quantum emitter inside the cavity leads to a new regime of cavity quantum electrodynamics, where the light--matter interaction dynamics is fundamentally different from that observed in a conventional cavity with Lorenztian lineshape. Furthermore, when the vibrational dynamics of the emitter is taken into account, an intricate phonon--photon interplay arises, and the optical interference induced by the Fano mirror significantly alters the leakage of energy into vibrational modes. We demonstrate that this control mechanism improves the maximum attainable indistinguishability of emitted photons, as compared to an equivalent cavity with a conventional mirror.
\end{abstract}
\maketitle

\begin{figure}
	\centering
	\includegraphics[width=\columnwidth]{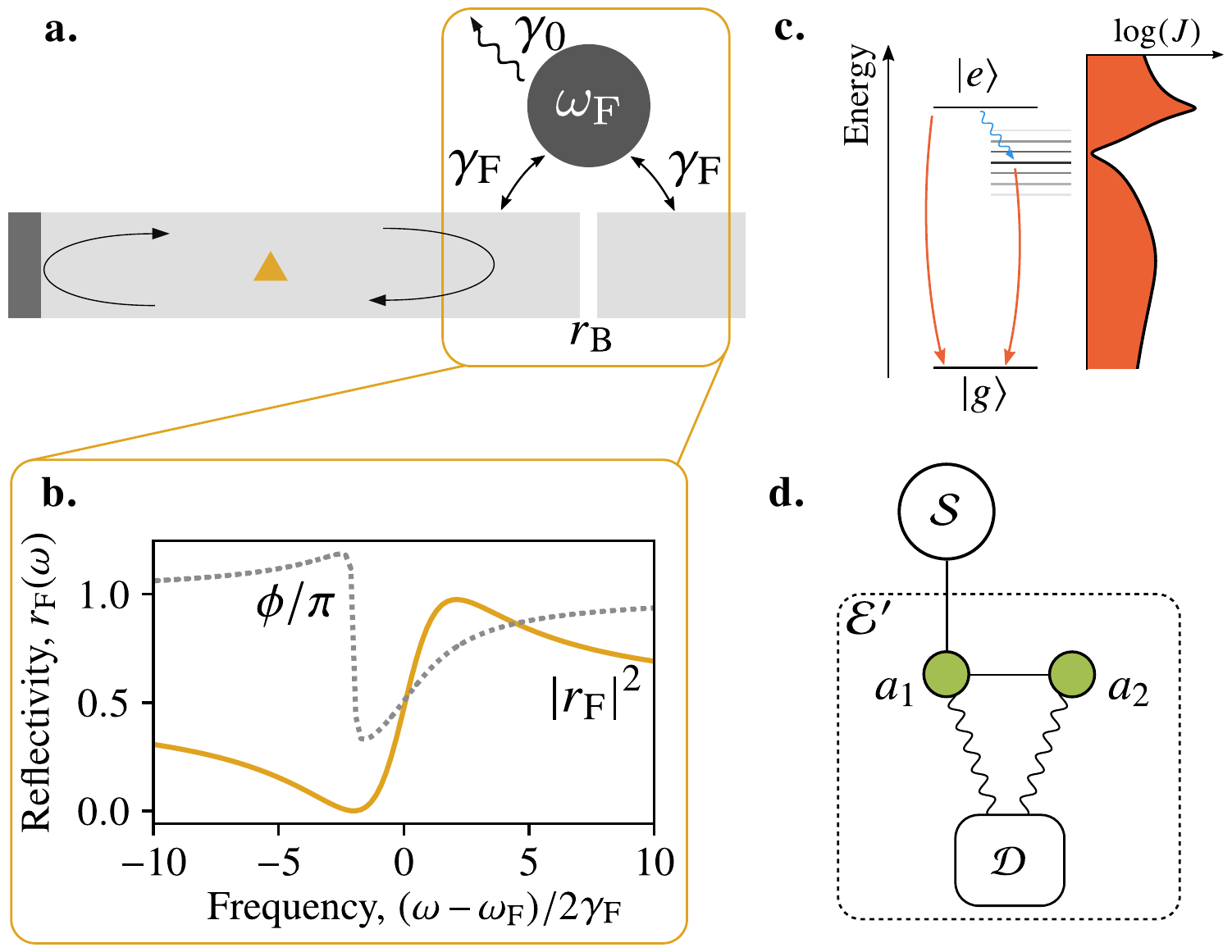}
	\caption{
{\bf a.} Fano cavity consisting of a waveguide with a fully reflecting mirror in the left end and a Fano mirror in the right end. 
{\bf b.} Suared modulus (orange solid) and phase (grey dotted) of the Fano mirror reflectivity, with $\gamma_0=0.05\gamma_\mathrm{F}$ and $r_\mathrm{B}=-1/\sqrt{2}$. 
{\bf c.} Level structure of quantum emitter ($\ket{e}$ and $\ket{g}$) and phonon modes along with optical local density of states, $J$.
{\bf d.} The photonic structure is mapped to a simple structure, $\mathcal{E}'$, consisting of two coupled discrete modes, dissipating into a common reservoir. 
}
	\label{fig:1}
\end{figure}

The Fano effect arises from the interference between a continuum of modes and a localised resonance~\cite{fano1961effects}. A photonic variant of the phenomenon can be realised by placing a nanocavity close to a waveguide with a partially transmitting element, as shown in Fig.~\ref{fig:1}a. In this configuration,  photons in the waveguide can propagate directly through the partially transmitting element or via the cavity. The resulting interference leads to a strong and asymmetric frequency dependence of the transmittivity through the system~\cite{heuck2013improved} (see Fig.~\ref{fig:1}b). 
By replacing one of the mirrors in a conventional Fabry-P\'erot cavity with such a Fano mirror, one may construct an exotic nanophotonic structure -- a so-called Fano cavity. 
The interference between the two dissipation paths of the Fano mirror leads to a quasi-localised cavity mode with an asymmetric line-shape that is far from a Lorentzian Fabry-P\'erot cavity. 
Indeed, in the limit where the two paths exhibit perfect destructive interference, this mode becomes a true bound state in the continuum~\cite{friedrich1985interfering}, which has recently been studied with a non-linear atomic mirror that allows for excitation of the bound mode through a two-photon process~\cite{cotrufo2018single,calajo2019exciting}.
Such Fano cavities are of growing interest in the field of integrated nanophotonics, where Fano lasers with unique dynamics~\cite{yu2017demonstration,mork2014photonic} and non-reciprocal elements~\cite{yu2015nonreciprocal} have been demonstrated.
However, there are a number of open questions regarding how the non-Lorentzian nature of the cavity will impact the emission properties of a single quantum emitter.

In this paper, we extend the description of a Fano cavity into the regime of cavity quantum electrodynamics (cQED), where we consider a single solid-state emitter placed in the centre of the cavity.
This theory allows us to directly investigate the impact that the non-Lorentzian lineshape of the Fano cavity has on the optical properties of the emitter, and crucially demonstrate that the coupling between a solid-state emitter and its vibrational environment can be engineered by structuring the optical local density of states (LDOS). This effect is observed through drastic modifications to the phonon sideband present in solid-state emitters~\cite{besombes2001acoustic}, which we show becomes highly sensitive to the shape of the LDOS of the nanophotonic structure, and is able to suppress phonon emission (see Fig.~\ref{fig:1}c), resulting in photon indistinguishabilities that surpass standard Lorentzian cavities. 

When studying light-matter coupling in cQED, quantisation of the cavity mode is of paramount importance. Therefore, a central question in the analysis of the Fano cavity is how to fully represent the structured LDOS in regimes of potentially strong light-matter coupling. Recently, a general quantisation scheme based on symmetrised quasi-normal modes has been presented, which relies on a numerical calculation of the electromagnetic modes of the system~\cite{franke2019quantization}. In contrast, we make use of a simple and intuitive approach that provides an analytical expression for the LDOS of the Fano cavity. 
We then develop a mapping that extracts the key features of the LDOS and incorporates them into an enlarged system Hamiltonian. 
This mapping generates a pair of dissipative bosonic modes (see Fig.~\ref{fig:1}d) which are coupled in order to capture the predominant interference pathways, and thus the non-Lorentzian behaviour of the Fano cavity. Resolving such interference features requires at least two coupled modes, and thus cannot be analysed with any single-mode mapping technique such as the reaction coordinate master equations~\cite{iles2014environmental,iles2016energy}.
 Our two-mode representation allows us to fully account for strong-coupling effects between the emitter and the Fano cavity, while simultaneously capturing the non-Markovian dynamics arising from interactions with the vibrational phonon environment of the host lattice~\cite{roy2015quantum,iles2017phonon}.

\begin{figure}
  \centering
  \includegraphics[width=\columnwidth]{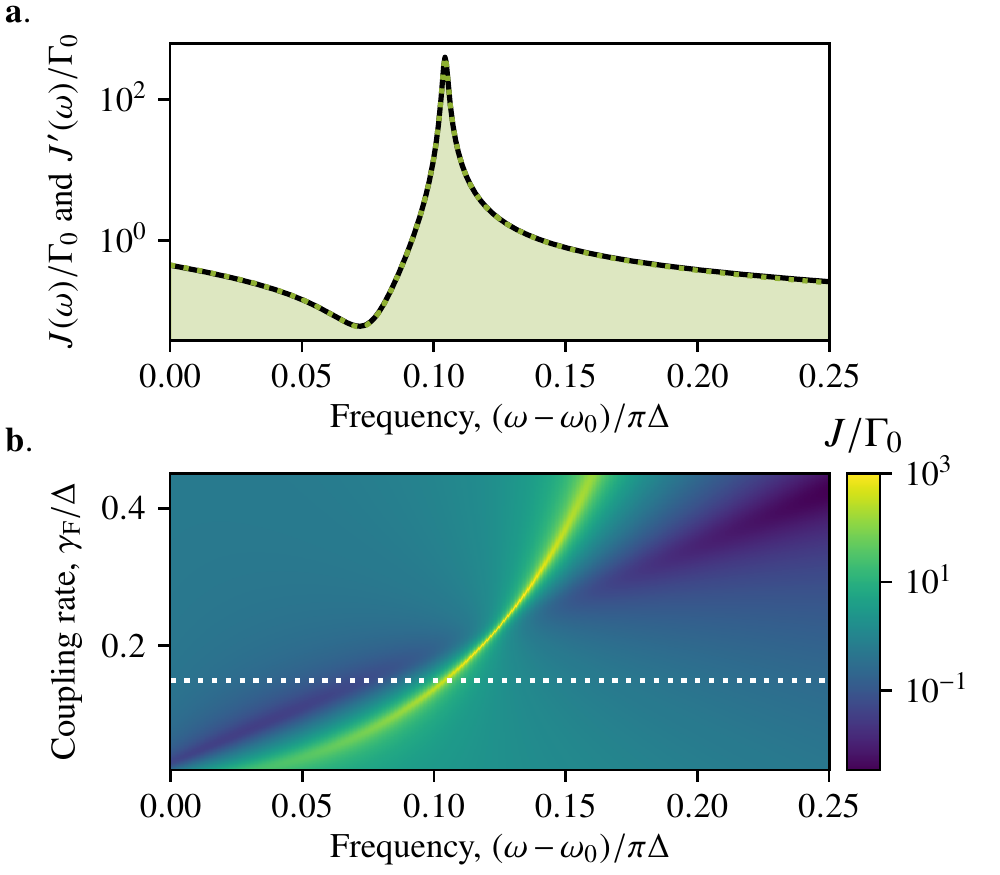}
  \caption{{\bf a.} Frequency dependence of the LDOS, $J/\Gamma_0$, (black solid) corresponding to Eq.~\eqref{eq:10} and spectral density, $J'/\Gamma_0$, generated by the mapped structure (green dotted line, shaded area). Parameters: $r_\mathrm{B}=-1/\sqrt{2},\; P=+1,\; \gamma_0=10^{-3}\Delta,\; \omega_0=101\pi\Delta,\;\omega_\mathrm{F}=\omega_0-0.02\pi\Delta,\; \gamma_\mathrm{F}=0.15\Delta$. {\bf b.} LDOS as a function of frequency and coupling rate. The white dotted line indicates the value of $\gamma_\mathrm{F}$ used in panel a, and the parameters are the same as in panel a.}
  \label{fig:2}
\end{figure}

In the rotating wave approximation, the Hamiltonian describing the interaction between a single two level emitter and the electromagnetic field is $H_\mathrm{EF}=\sum_\mu (g_\mu a_\mu^\dagger\sigma + g_\mu^*a_\mu \sigma^\dagger)$, where $a_\mu$ is the bosonic annihilation operator for the $\mu^{\rm th}$-mode of the field, $\sigma=\dyad{g}{e}$ is the emitter transition operator between the excited ($\ket{e}$) and ground ($\ket{g}$) states, and $g_\mu$ is the coupling constant. The spectral density, $J(\omega)=2\pi\sum_\mu \abs{g_\mu}^2\delta(\omega-\omega_\mu)$, with $\omega_\mu$ the frequency of the $\mu^{\text{th}}$ mode, then fully characterises the interaction. This spectral density is equivalent to the LDOS of the electromagnetic field~\cite{lodahl2015interfacing}, which can be calculated using an extension of Refs.~\cite{gregersen2016broadband,rybin2016purcell,denning2018cavity} where the right mirror is a Fano mirror with reflectivity $r_\mathrm{F}(\omega)=  r_\mathrm{B} + [-i(\omega-\omega_\mathrm{F})/2\gamma_\mathrm{F}+\gamma_\mathrm{t}/2\gamma_\mathrm{F}]^{-1}[-r_\mathrm{B}+iPt_\mathrm{B}]$~\cite{yoon2013fano,heuck2013improved,yu2015nonreciprocal}. Here, $r_\mathrm{B}$ is the bare reflectivity of the partially transmitting element in the waveguide; $\gamma_\mathrm{F}$ is the coupling rate between the nanocavity and the waveguide, which is taken equal on both sides of the cavity; $\gamma_0$ is the intrinsic loss rate of the nanocavity; $\gamma_\mathrm{t}=2\gamma_\mathrm{F}+\gamma_0$ is the total loss rate; $\omega_\mathrm{F}$ is the resonant frequency of the nanocavity and $P$ is the parity ($\pm 1$) of the Fano mirror, determining whether the reflectivity minimum is on the blue- or red-detuned side of the maximum~\cite{osterkryger2016spectral,heuck2013improved}. When the emitter is placed in the middle of the cavity, the LDOS becomes~\cite{suppmat}
\begin{align}
  \label{eq:10} J(\omega)=\Gamma_0\Re\qty[\frac{(1+r_0e^{i\omega/\Delta)})(1+r_\mathrm{F}(\omega)e^{i\omega/\Delta})}{1-r_0r_\mathrm{F}(\omega)e^{2i\omega/\Delta}}],
\end{align}
where $\Gamma_0$ is the spontaneous emission into the bare waveguide in the absence of mirrors, $r_0$ is the reflectivity of the conventional left mirror (here taken to be $-1$) and $\Delta/2$ is the free spectral range of the bare resonator set by $\Delta=c/(\bar{n}L)$, with $c$ the speed of light, $\bar{n}$ the effective waveguide group index and $L$ the cavity length. 
The interplay between the roundtrip phase in the cavity and the reflection phase of the Fano mirror gives rise to a resonance and an anti-resonance appearing as a peak and a dip in the LDOS (see Fig.~\ref{fig:2}a). Fig.~\ref{fig:2}b shows how the LDOS depends on frequency and the coupling rate between the nanocavity and the waveguide. 
At the critical point where the anti-resonance crosses over from the red-detuned to the blue-detuned side of the resonance, the peak corresponds to a bound state in the continuum~\cite{friedrich1985interfering}. At this point,
the lifetime of this bound state is only limited by the intrinsic losses of the nanocavity.

Various techniques have been developed to perform the non-trivial task of constructing an accurate representation of a continuous bosonic environment in terms of a set of discrete modes. Reaction coordinate master equations have previously been used to extract a single damped bosonic mode from a continuous reservoir~\cite{iles2014environmental,iles2016energy}. Such a representation, however, has so far only been successful in representing environments that are Lorentzian in nature, and therefore cannot capture the central features of the Fano cavity induced by interference. The pseudomode mapping associates a discrete mode with each complex pole of the spectral density~\cite{garraway1997nonperturbative,dalton2001theory,mazzola2009pseudomodes}. This technique is exact but requires that all poles in the spectral density are included in the mapping, because the sum of the residues must have a vanishing real part. In the present paper, the spectral density has infinitely many poles, and therefore such an approach is infeasible. 
A more flexible approach allows representation of a bosonic environment as an infinite discrete chain of coupled harmonic oscillators~\cite{chin2010exact}, with the possibility of truncating the chain after a finite number of links~\cite{woods2014mappings}. 
In order to be efficient, the chain needs to represent the spectral density accurately with only a few links, or rely on sophisticated numerical techniques to calculate the dynamics of the reduced system~\cite{Prior2010,Rosenbach_2016}.
However, we have found that for the Fano cavity, no clear convergence within 10 chain sites is observable. This is largely attributed to the interference between the two optical decay paths in the physical structure that generates the central features of the spectral density, which is challenging to reproduce in an equivalent one-dimensional system.

On a more general note, it was recently shown that the reduced dynamics of a quantum system, $\mathcal{S}$, with a bosonic environment, $\mathcal{E}$, is invariant under a substitution of $\mathcal{E}$ by a different bosonic environment, $\mathcal{E}^\prime$, provided that the spectral densities of $\mathcal{E}$ and $\mathcal{E}^\prime$ are equal~\cite{tamascelli2018nonperturbative}. In most practical applications, the correspondence between the spectral densities of $\mathcal{E}$ and $\mathcal{E}^\prime$ cannot be exact, but their difference within a relevant frequency range determined by $\mathcal{S}$ serves as a measure for equivalence of a given mapping. This strategy can be exemplified by the canonical representation of a QED emitter--cavity system by the dissipative Jaynes-Cummings model, where the cavity is described by a single mode with a finite lifetime. To describe the asymmetric Fano cavity mode, it is necessary to use two modes (see Fig.~\ref{fig:1}d), with annihilation operators $A=(a_1,a_2)^T$ where light-matter coupling is described by the master equation
\begin{align}
  \label{eq:6}
  \dot{\rho}=-i[H_\mathrm{E} + A^\dagger\Omega A + A^\dagger G\sigma + G^\dagger A \sigma^\dagger,\rho(t)]+\mathcal{D}(K^\dagger A).
\end{align}
Here, $H_\mathrm{E}=\omega_{eg}\sigma^\dagger\sigma$ is the free evolution Hamiltonian of the emitter, $\mathcal{D}(x)=x\rho(t) x^\dagger - \frac{1}{2}(x^\dagger x\rho(t)+\rho(t) x^\dagger x)$ is the Lindblad dissipator and
\begin{align}
  \label{eq:8}
\Omega=\mqty[\omega_1&V\\V^*&\omega_2], \; G=\mqty[g\\0], \; K=\mqty[\sqrt{\kappa_1}\\ \sqrt{\kappa_2}].
\end{align}
We have introduced the emitter-cavity coupling strength $g$, the resonant frequency $\omega_i$ and leakage rate $\kappa_i$ for the $i^{\rm th}$-cavity mode, and the intercavity coupling strength $V$.
The interference between the two modes appearing in the dissipator is crucial for capturing the underlying physical interference between the direct and cavity-mediated transmissions in the Fano mirror.

To establish a link between the microscopic physical parameters of the cavity and the parameters in the discrete representation, we calculate the spectral density generated by the $A$-modes and compare it to the LDOS of Eq.~\eqref{eq:10} to minimise the error, $\epsilon=\int_W\dd{\omega}[J(\omega)-J'(\omega)]^2$, over a finite frequency window, $W$, around the emitter resonance. The spectral density of the discrete mode representation is calculated as $J'(\omega)=g^2\int_{-\infty}^\infty\dd{\tau}\ev*{a_1(\tau)a_1^\dagger}e^{i\omega\tau}$, where $\ev*{\cdot}$ denotes the free evolution of the cavity modes under Eq.~\eqref{eq:6} in the absence of the emitter. With analytic expressions for $J$ and $J'$, we require that the position of the complex poles of $J'$ is equal to those of $J$, and the imaginary part of the sum of the residues are equal~\cite{suppmat}. These requirements determine five out of the seven parameters in the discrete-mode representation. The remaining two parameters are determined numerically by minimising $\epsilon$.
The LDOS in Eq.~\eqref{eq:10} features an infinite number of discrete poles corresponding to the Fabry-Pérot resonances. However, the emitter is only sensitive to the the LDOS in the vicinity of its resonance frequency. For cavities with length scales of a few micrometers, the free spectral range is much larger than any features in the spectral response of the emitter and phonon modes, covering a few meV. As such, we find that features of the bound mode in the Fano cavity are very well described by two nearby poles of $J$, separated from the remaining poles by $\sim \pi\Delta$. In Fig.~\ref{fig:2}a, we provide an example of how the mapped spectral density, $J'$, (green dashed line and shaded area) reproduces with a high precision the LDOS, $J$ (black solid line). A recent related work presents a similar technique based on a numerical search for the optimal parameters for a coupled network of harmonic oscillators to represent a given spectral density~\cite{mascherpa2019optimized}. 

\begin{figure}
  \centering
  \includegraphics[width=\columnwidth]{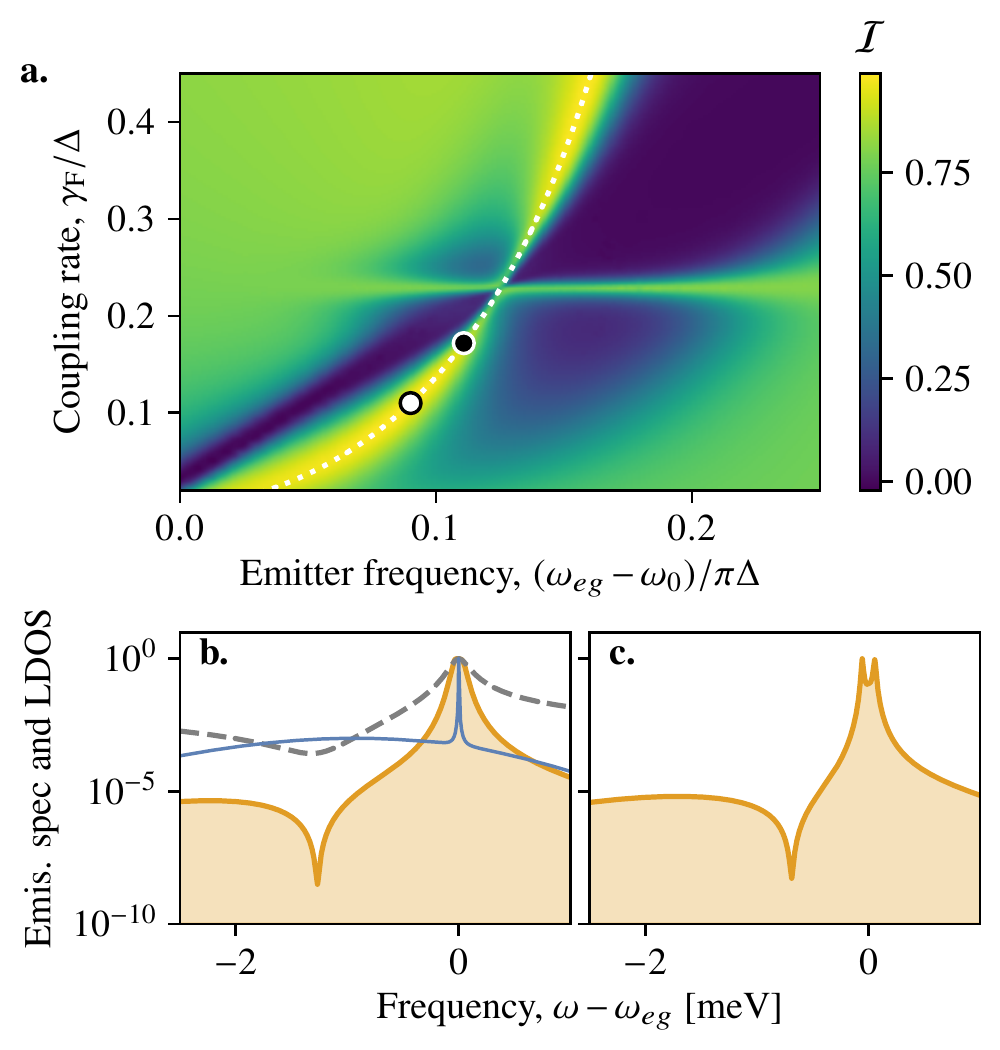}
  \caption{{\bf a.} Photon indistinguishability (colour scale) for varying emitter frequency, $\omega_{eg}$, and nanocavity coupling rate, $\gamma_\mathrm{F}$. The dotted white line traces the peak of the LDOS in Fig.~\ref{fig:2}b. Parameters: $\Delta=10\mathrm{\;meV},\; \Gamma_0=0.6\mathrm{\; \mu eV},\; \Gamma_\mathrm{R}=0.03\mathrm{\: \mu eV},\; \alpha=0.069\mathrm{\; meV^{-2}},\; \nu_c=1.45\mathrm{\: meV},\; T=4\mathrm{\: K}$ and otherwise as in Fig.~\ref{fig:2}. {\bf b.} Emission spectrum (solid orange and shaded area) for optimal parameters with $\delta=0.010$ corresponding to the white dot in panel a along with the LDOS (grey dashed) and the bulk emission spectrum. All quantities are normalised to their peak value. {\bf c.} Emission spectrum for parameters corresponding to the black dot in panel a. }
  \label{fig:3}
\end{figure}

Interactions with longitudinal acoustic phonons are described by the additional Hamiltonians $H_\mathrm{P}=\sum_\mathbf{q}\nu_\mathbf{q}b_\mathbf{q}^\dagger b_\mathbf{q}$ and $H_\mathrm{EP}=\sigma^\dagger \sigma\sum_\mathbf{q} M_\mathbf{q} (b_\mathbf{q}^\dagger+b_\mathbf{q})$~\cite{wilson2002quantum,mahan2013many,kaer2014decoherence}. The phonon interactions are characterised by the spectral density $J_\mathrm{P}(\nu)=\sum_\mathbf{q}M_\mathbf{q}^2\delta(\nu-\nu_\mathbf{q})=\alpha \nu^3 \exp[-\nu^2/\nu_c^2]$, with overall coupling strength $\alpha$ and cutoff frequency $\nu_c$~\cite{nazir2008photon,nazir2016modelling}. To account for the non-Markovian emitter--phonon correlations, we use the polaron transformation ~\cite{wilson2002quantum,iles2017phonon,nazir2016modelling,mccutcheon2010quantum,roy2015quantum} to derive a master equation~\cite{suppmat}. The resulting master equation is a modification of Eq.~\eqref{eq:6} in which the light--matter coupling rate, $g$, is effectively reduced and an additional dissipator is introduced. We furthermore add a dissipator accounting for spontaneous emission into the radiation modes, $\Gamma_\mathrm{R}L(\sigma)$~\cite{mccutcheon2013model,roy2015spontaneous}.


The photon indistinguishability is an important feature that characterises the coherence properties of the light emitted from a cQED system~\cite{iles2017phonon,roy2015quantum,kaer2010non,kaer2014decoherence,Grange2017green}. In the present situation, the indistinguishability reveals how strongly the photonic structure is able to modify the phonon dynamics by shaping the phonon sideband or enhancing/suppressing the zero phonon line. Here, the emission from the system is studied by choosing an initial state where the emitter is excited and the electromagnetic field is in the vacuum state, and then calculating the optical spectrum as the system relaxes and a single photon is emitted. Of particular interest is the two-colour spectrum of light emitted from the system through the Fano mirror, $S(\omega,\omega')$~\cite{iles2017phonon}, which is related to the dipole spectrum, $S_0(\omega,\omega')=\int_{-\infty}^\infty\dd{t}\dd{t'}e^{i(\omega t-\omega't')}\ev*{\sigma^\dagger(t)\sigma(t')}$, through the optical Green's function as $S(\omega)=G^*(\omega)G(\omega')S_0(\omega,\omega')$, where $G(\omega)=[1+r_0e^{i\omega/\Delta}]t_\mathrm{F}(\omega)[1-r_0r_\mathrm{F}(\omega)e^{2i\omega/\Delta}]^{-1}$ and $t_\mathrm{F}$ is the transmittivity of the Fano mirror. This two-colour spectrum is related to the observable emission spectrum, $\bar{S}(\omega)$, as $\bar{S}(\omega)=S(\omega,\omega)$. Furthermore, it allows calculation of photon indistinguishability as~\cite{suppmat,iles2017phonon,denning2018cavity}
\begin{align}
  \label{eq:1}
  \mathcal{I}=[2P/\Gamma_0]^{-2}\int_{-\infty}^\infty\dd{\omega}\dd{\omega'}\abs{S(\omega,\omega')}^2,
\end{align}
where $P=(\Gamma_0/2)\int_{-\infty}^\infty\dd{\omega}\bar{S}(\omega)$ is the emitted energy into the detection channel. 
In a multiphoton experiment with $n$ subsequent two-photon interference events, the accumulated visibility scales as $\mathcal{I}^n \simeq 1-n\delta$, where $\delta:= 1-\mathcal{I}$ is the distinguishibility, which quantifies the decoherence-induced error~\cite{kiraz2004quantum}.

Fig.~\ref{fig:3}a shows the photon indistinguishability as a function of nanocavity coupling rate and emitter transition frequency, corresponding to parameter variations over the LDOS shown in Fig.~\ref{fig:2}b. The behaviour of the indistinguishability over the parameter space shows a rich structure, which is mainly a result of the interplay between the electromagnetic LDOS and the phonon sideband. Along the dotted line, the emitter transition frequency is tuned to the resonance of the Fano cavity, meaning that emission is funnelled into the zero-phonon line. 
In the lower part of the plot, the LDOS anti-resonance is located on the red-detuned side of the resonance and thereby suppresses emission into the phonon sideband while enhancing emission into the zero phonon line. 
The combination of these effects leads to a minimal distinguishability of 0.010 (indicated with a white dot). 
The emission spectrum corresponding to these parameters is shown in Fig.~\ref{fig:3}b (solid orange; shading). 
To understand how this spectrum arises, the bulk emission spectrum is indicated (thin blue) along with the electromagnetic LDOS (grey dashed). This demonstrates how the anti-resonance leads to a spectral hole in the phonon sideband. 
In the supplemental material~\cite{suppmat}, we show that the minimal attainable distinguishability for the same optical structure with a conventional mirror is $\delta=0.013$~\cite{iles2017phonon,denning2018cavity}, found when the mirror reflectivity is optimised to $r\simeq 0.99$. 
Fig.~\ref{fig:3}c shows the emission spectrum for parameters corresponding to the black dot in Fig.~\ref{fig:3}a, where the cavity lifetime is long enough for the system to enter the strong coupling regime and two polariton peaks can be seen along with the spectral hole in the phonon sideband. In this regime, phonons reduce the indistinguishability by driving transitions between the polaritons~\cite{kaer2010non,kaer2014decoherence,iles2017phonon}. 
If the emitter frequency is blue-shifted relative to the resonance line in Fig.~\ref{fig:3}a, the LDOS peak will lie in the sideband, thereby increasing the phonon sideband emission, followed by a reduction of indistinguishability. Similarly, the low-indistinguishability area in Fig.~\ref{fig:3}a on the blue-detuned side of the resonant line arises because the emitter transition lies on the cavity anti-resonance, thereby suppressing emission into the zero-phonon line.

In conclusion, we have analysed the quantum light-matter interactions between a single emitter and an optical Fano cavity. 
To this end, we have developed a mapping technique that allows to accurately represent the non-Lorentzian features of the cavity. 
We have demonstrated that coupling to the Fano cavity leads to rich and complex optical behaviour, and in particular shown how the anti-resonance of the Fano cavity can shape the phonon sideband, thereby increasing the photon indistinguishability to values unobtainable with conventional Fabry-P\'erot cavities.

During the preparation of this manuscript, we became aware of related work~\cite{vcernotik2019cavity} that presents a quantum optical description of the Fano mirror.

E.V.D. and J.M. acknowledge funding from the Danish Council for Independent Research (Grant No. DFF-4181-00416) and from Villum Fonden (Grant No. 8692). 
J.I.-S. acknowledges support from the Royal Commission for the Exhibition of 1851.

\clearpage

\setcounter{page}{1}
\setcounter{figure}{0}
\setcounter{equation}{0}
\renewcommand{\theequation}{S\arabic{equation}}
\renewcommand{\thefigure}{S\arabic{figure}}
\renewcommand{\bibnumfmt}[1]{[S#1]}
\renewcommand{\citenumfont}[1]{S#1}
\begin{center}
\textbf{\Large Supplemental Material}\\
Here we elaborate on the expressions and results presented in the main text.
\end{center}

\section{Reflectivity and transmittivity of Fano mirror}
\label{sec:refl-transm-fano}
\begin{figure}
  \centering
  \includegraphics[width=\columnwidth]{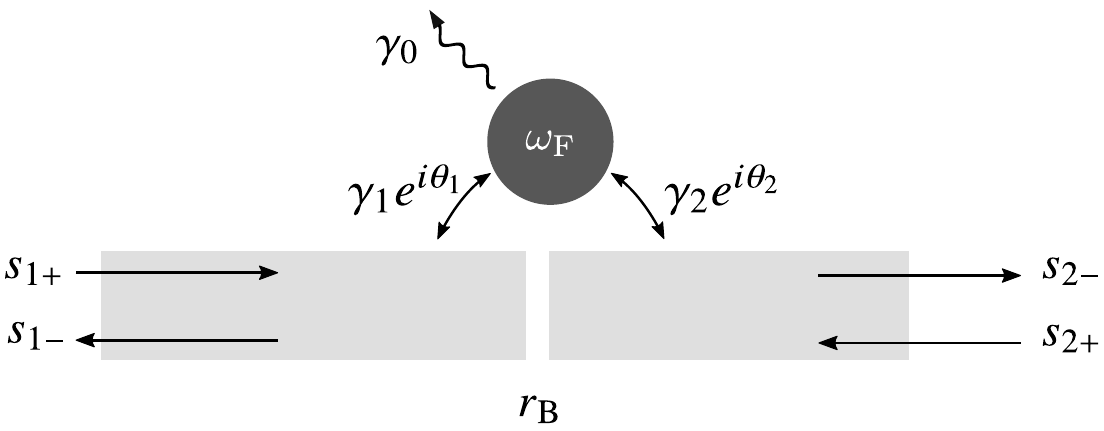}
  \caption{Fano mirror with in- and outgoing field amplitudes $s_i$ indicated. }
  \label{si:fig:fano-resonance}
\end{figure}

Here, we calculate the reflectivity and transmittivity of the bare Fano mirror for a general situation, where the side-coupled nanocavity couples to the semi-infinite left and right waveguide segments with different rates, $\gamma_1,\;\gamma_2$, and corresponding phases, $\theta_1,\theta_2$, (see Fig. \ref{si:fig:fano-resonance}). The reflectivity and transmittivity of the partially transmitting element are denoted by $r_\mathrm{B}$ and $t_\mathrm{B}$ and the intrinsic loss rate of the nanocavity is $\gamma_0$.
	
To analyse this system, we use coupled mode theory as in Refs.~\cite{si:suh2004temporal,si:yoon2013fano,si:wang2013fundamental,si:kristensen2017theory}, for the for the four in- and outgoing field amplitudes in the waveguide, $s_i$, as indicated in Fig.~\ref{si:fig:fano-resonance}, and in the resonant mode of the nanocavity, $a$,
\begin{align}
  \begin{split}
    \dv{a}{t}&=(-i\omega_\mathrm{F}-\gamma_t)a + \mqty(d_1 & d_2)\mqty(s_{1+}\\s_{2+}), \\
    \mqty(s_{1-}\\s_{2-}) &= C\mqty(s_{1+}\\s_{2+}) +  \mqty(d_1 \\ d_2)a, \;\; C=\mqty(r_\mathrm{B}&-it_\mathrm{B}\\-it_\mathrm{B}&r_\mathrm{B}),
  \end{split}
\end{align}
with $\gamma_t=\sum_i \gamma_i$, $d_i=\sqrt{2\gamma_i}e^{i\theta_i}$ and $\omega_\mathrm{F}$ the resonance frequency of the cavity.  
Fourier transforming the equations, setting $s_{2+}=0$ and defining the amplitude reflectivity and transmittivity as $r_\mathrm{F}(\omega)=s_{1-}/s_{1+},\; t_\mathrm{F}(\omega)=s_{2-}/s_{1+}$, we find
\begin{align}
  \begin{split}
    \label{si:eq:Fano-r-t-1}
    r_\mathrm{F}(\omega) &= r_\mathrm{B} + \frac{d_1^2}{-i(\omega-\omega_\mathrm{F}) + \gamma_t}, \\
    t_\mathrm{F}(\omega) &= -it_\mathrm{B} + \frac{d_1d_2}{-i(\omega-\omega_\mathrm{F}) + \gamma_t}.
  \end{split}
\end{align}
Using the relation $D=-CD^*$~\cite{si:suh2004temporal}, with 
\begin{align}
  D=\mqty(d_1&0\\d_2&0),
\end{align}
we arrive at the expressions
\begin{align}
  \label{si:eq:Fano-r-t-2}
  r_\mathrm{F}(\omega)&=r_\mathrm{B} + 2\frac{it_\mathrm{B}\sqrt{\gamma_1\gamma_2}e^{i(\theta_1-\theta_2)}-r_\mathrm{B}\gamma_1}{-i(\omega-\omega_\mathrm{F}) + \gamma_t}, \\
  t_\mathrm{F}(\omega)&= -it_\mathrm{B} + 2\frac{it_\mathrm{B}\gamma_2 - r_\mathrm{B}\sqrt{\gamma_1\gamma_2}e^{i(\theta_2-\theta_1)}}{-i(\omega-\omega_\mathrm{F})+\gamma_t}.
\end{align}
The phases are not independent from the coupling parameters, but are given as~\cite{si:yu2015nonreciprocal,si:yu2017demonstration}
\begin{align}
  \label{si:eq:phase-specifications}
  \begin{split}
    \cos 2\theta_1 &=\frac{1}{2}\frac{t_\mathrm{B}^2}{r_\mathrm{B}}\qty(\frac{\gamma_2}{\gamma_1}-1) - r_\mathrm{B}, \\
    \sin 2\theta_1 &=P t_\mathrm{B} \frac{\sqrt{4\gamma_1\gamma_2 - t_\mathrm{B}^2(\gamma_1+\gamma_2)^2}}{2\gamma_1 r_\mathrm{B}}, \\
    e^{i(\theta_1-\theta_2)} &=\sqrt{\gamma_1/\gamma_2}\frac{1}{it_\mathrm{B}}\qty(e^{2i\theta_1} +r_\mathrm{B})
  \end{split}
\end{align}
with $P$ the parity of the resonance, determined by the symmetry properties of the optical mode in the side-coupled cavity~\cite{si:osterkryger2016spectral}.

\subsection{Symmetric couplings}
\label{sec:sym-coupling}
If we now assume that the couplings are symmetric, $\gamma_1=\gamma_2 =: \gamma_\mathrm{F}$, we can reduce Eqs.~\eqref{si:eq:Fano-r-t-1} and \eqref{si:eq:Fano-r-t-2} to 
\begin{align}
  \begin{split}
    r_\mathrm{F}(\omega) &= r_\mathrm{B} + \frac{2\gamma_\mathrm{F} e^{2i\theta_1}}{-i(\omega-\omega_\mathrm{F}) + \gamma_t}, \\
    t_\mathrm{F}(\omega) &= -it_\mathrm{B} + \frac{2\gamma_\mathrm{F}(it_\mathrm{B}-r_\mathrm{B}e^{i(\theta_2-\theta_1)}) }{-i(\omega-\omega_\mathrm{F}) + \gamma_t}.
  \end{split}
\end{align}
These expressions can be further simplified by using the relations Eq.~\eqref{si:eq:phase-specifications}, where we get $\cos 2\theta_1=-r_\mathrm{B},\; \sin 2\theta_1 = P t_\mathrm{B},\; e^{i(\theta_1-\theta_2)}=P$. The reflectivity and transmittivity then become
\begin{align}
  \label{si:eq:Fano-r-t-symmetric}
  \begin{split}
    r_\mathrm{F}(\omega) &= r_\mathrm{B} + \frac{-r_\mathrm{B} + iPt_\mathrm{B}}{-i(\omega-\omega_\mathrm{F})/2\gamma_\mathrm{F} + \gamma_t/2\gamma_\mathrm{F}}, \\
    t_\mathrm{F}(\omega) &= -it_\mathrm{B} + \frac{it_\mathrm{B} - P r_\mathrm{B}}{-i(\omega-\omega_\mathrm{F})/2\gamma_\mathrm{F} + \gamma_t/2\gamma_\mathrm{F}}.
  \end{split}
\end{align}

\subsection{Asymmetric couplings}
\label{sec:asym-coupling}
In the general case, where the couplings are asymmetric, we define the asymmetry ratio, $\chi=\gamma_2/\gamma_1$. With this, the phase relations in Eq.~\eqref{si:eq:phase-specifications} become
\begin{align}
  \label{si:eq:phase-specification-chi}
  \begin{split}
    \cos 2\theta_1 &= \frac{1}{2}\frac{t_\mathrm{B}^2}{r_\mathrm{B}} (\chi-1) - r_\mathrm{B}, \\
    \sin 2\theta_1 &= \frac{P t_\mathrm{B}\sqrt{4\chi-t_\mathrm{B}^2(1+\chi)^2}}{2r_\mathrm{B}}, \\
    e^{i(\theta_1-\theta_2)} &= \sqrt{1/\chi} \frac{1}{it_\mathrm{B}}(e^{2i\theta_1}+r_\mathrm{B}) \\
    = \sqrt{1/\chi}\frac{1}{2ir_\mathrm{B}}&\qty(t_\mathrm{B}(\chi-1)+iP\sqrt{4\chi-t_\mathrm{B}^2(1+\chi)^2}),
  \end{split}
\end{align}
leading to 
\begin{align}
  \label{si:eq:Fano-r-t-asymmetric}
  \begin{split}
    r_\mathrm{F}(\omega) &= r_\mathrm{B} \\ &+ \frac{\frac{t_\mathrm{B}}{2r_\mathrm{B}}\qty[t_\mathrm{B}(\chi-1) + iP\sqrt{4\chi-t_\mathrm{B}^2(1+\chi)^2}]-r_\mathrm{B}}{-i(\omega-\omega_0)/2\gamma_1 + \gamma_t/2\gamma_1}, \\
    t_\mathrm{F}(\omega) &= -it_\mathrm{B} \\ &+\frac{it_\mathrm{B} - \frac{1}{2} \qty{it_\mathrm{B}(\chi-1) + P \qty[\sqrt{4\chi-t_\mathrm{B}^2(1+\chi)^2}]^* }}{-i(\omega-\omega_0)/2\gamma_1 + \gamma_t/2\gamma_1}.
  \end{split}
\end{align}
As expected, the expressions reduce to Eq.~\eqref{si:eq:Fano-r-t-symmetric} in the symmetric limit $\chi = 1$.

\section{Local density of states of electromagnetic field in Fano cavity}
\label{sec:local-density-states}

\begin{figure}
  \centering
  \includegraphics[width=\columnwidth]{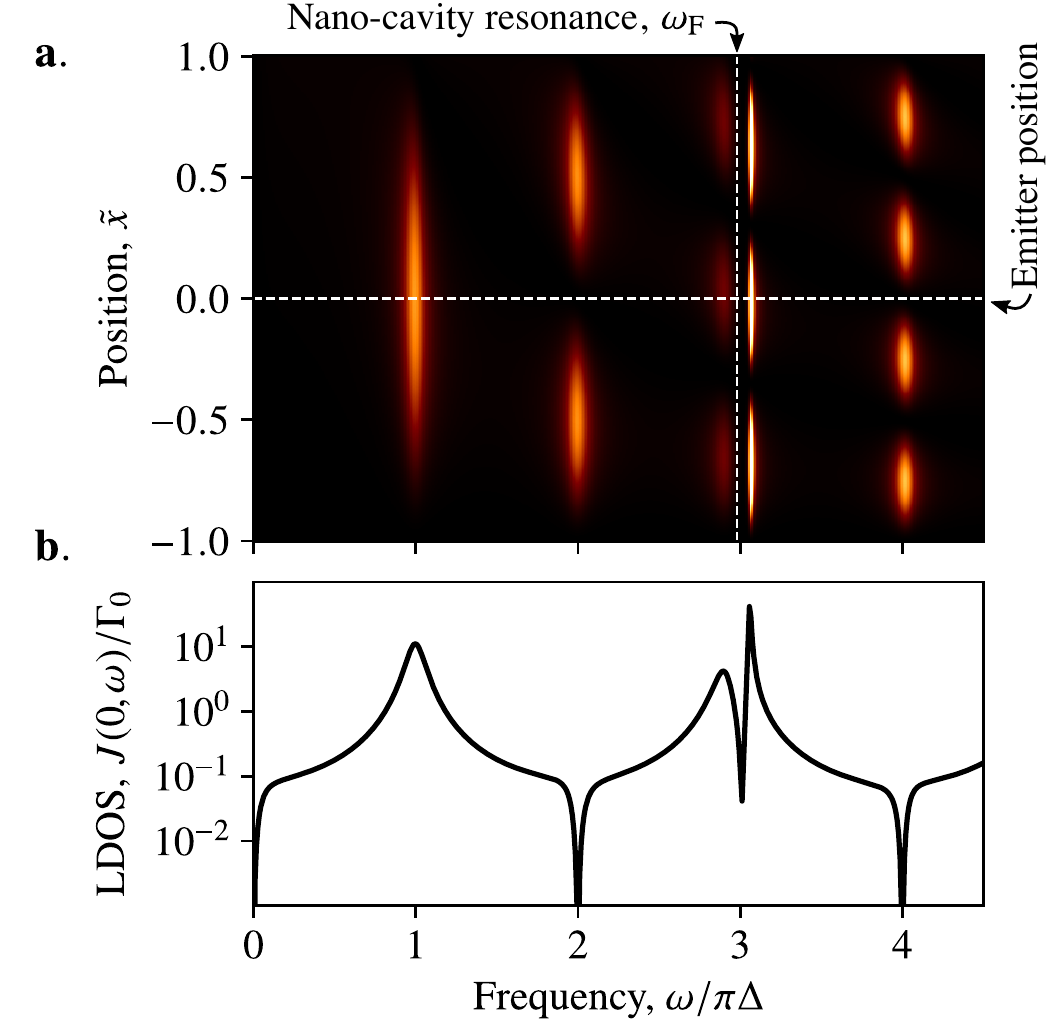}
  \caption{{\bf a.} LDOS of the Fano cavity as a function of position, $\tilde{x}$, and frequency, $\omega$, normalised to the free spectral range, $\Delta$. The resonance frequency of the side-coupled nanocavity and the emitter position are indicated with vertical and horizontal dashed lines, respectively. {\bf b.} LDOS as a function of frequency taken at the position in the middle of the cavity region, $\tilde{x}=0$. The parameters for both panels are $r_\mathrm{B}=-1/\sqrt{2},\; \gamma_1=\gamma_2=0.05\Delta,\; P=+1,\; \omega_F=2.98\Delta,\; \gamma_0=10^{-3}\Delta$.}
  \label{si:fig:ldos}
\end{figure}

Following Refs.~\cite{si:gregersen2016broadband,si:denning2018cavity} we consider a waveguide supporting a single guided transverse mode, with two mirrors forming a cavity. In the region between the two mirrors, the local density of states (LDOS) of the electromagnetic field associated with the cavity mode is
\begin{align}  J(x,\omega)=\Gamma_0\Re{\frac{[1+\tilde{r}_1(x,\omega)][1+\tilde{r}_2(x,\omega)]}{1-\tilde{r}_1(x,\omega)\tilde{r}_2(x,\omega)}},
\end{align}
where $x$ is the longitudinal position ($x=\pm L/2$ corresponding to the position of the two mirrors, respectively), $\tilde{r}_1(x,\omega)$ and $\tilde{r}_2(x,\omega)$ are the complex reflectivities of the left and right mirror in the cavity, respectively. These are defined such that $\tilde{r}_i(x,\omega)$ also accounts for the propagation from the position of the emitter, $x$, to the $i$'th mirror and back to $x$. Taking the left mirror (index 1) to be perfectly reflecting with a reflectance phase $\phi_1$ and the right mirror (index 2) to be a Fano resonance with reflectivity $r_\mathrm{F}(\omega)$,
\begin{align}
  \label{si:eq:5}
\begin{split}
  \tilde{r}_1(x,\omega)&=\exp{i[(1+\tilde{x})\omega \bar{n}L/c+\phi_1]} \\
\tilde{r}_2(x,\omega)&=r_\mathrm{F}(\omega)\exp{i[(1-\tilde{x})\omega \bar{n}L/c]},
\end{split}
\end{align}
where $c$ is the speed of light and  $\tilde{x}=2x/L$ is a dimensionless position along the waveguide axis, such that $\tilde{x}=-1$ corresponds to the position of the left mirror and $\tilde{x}=+1$ of the right mirror. The free spectral range (FSR) of the resonator is then given by $\Delta/2=c/(2\bar{n}L)$ (such that the corresponding FSR for angular frequencies, $\omega$, is $\pi\Delta$).

Fig.~\ref{si:fig:ldos}a shows the LDOS of the Fano-cavity, $J(\tilde{x},\omega)$, as a function of position and frequency. Far away from the resonance of the side-coupled cavity (indicated by a dashed vertical line) the partially transmitting element and the left mirror generate simple Fabry-Perot resonances separated by the free spectral range, $\pi\Delta$. Around the resonance of the side-coupled cavity, Fano interference effects give rise to a sharp dip in the LDOS located close to sharp peak. This is more clearly seen in Fig.~\ref{si:fig:ldos}b, where the LDOS for the cavity midpoint ($\tilde{x}=0$) is plotted.

 \section{Mapping}
\label{sec:mapping}

\subsection{Effective spectral density of few-mode reservoir}
\label{sec:effect-spectr-dens}

The full Hamiltonian for the solid state emitter and the electromagnetic field is 
\begin{align}
  \label{si:eq:102}
  H= H_\mathrm{E} + \sum_\mu \omega_\mu a_\mu^\dagger a_\mu + [g_\mu a_\mu^\dagger \sigma +g_\mu^* a_\mu\sigma^\dagger],
\end{align}
where $H_\mathrm{E}$ accounts for the free evolution of the emitter as well as the vibrational dynamics of the lattice.
The mapped environment consists of two coupled cavity modes, described by bosonic annihilation operators $a_1$ and $a_2$, both coupled to a common auxiliary bosonic reservoir (labeled by the symbol $\xi$) with annihilation operators $\alpha_\mu$. The emitter couples to the mapped environment only through the $a_1$ cavity mode (with the real coupling rate $g$). The full Hamiltonian for this system, $H'$, is
\begin{align}
  \label{si:eq:99}
\begin{split}
  H' &=H_\mathrm{E} + \sum_i\omega_ia_i^\dagger a_i+  [g a_1\sigma^\dagger + V a_1^\dagger a_2 + \mathrm{H.c.}]\\ &+ [q_1a_1^\dagger + q_2 a_2^\dagger]\sum_\mu f_\mu \alpha_\mu + [q_1^*a_1 + q_2^* a_2]\sum_\mu f_\mu^* \alpha_\mu^\dagger \\ &+ \sum_\mu \epsilon_\mu \alpha_\mu^\dagger \alpha_\mu
\end{split}
\end{align}
where $\epsilon_\mu$ and $f_\mu$ are the frequency and overall coupling rate to the $\mu^\text{th}$ mode in the auxiliary reservoir, $q_i$ are unitless complex numbers specifying the relative strength and phase of the mode--reservoir couplings, $V$ is a complex coupling rate between the two cavity modes and $\omega_1,\omega_2$ are their frequencies.

The light-matter coupling terms in both $H$ and $H'$ are of the form $\sigma c^\dagger+\sigma^\dagger c$, where for $H$ we have $c=\sum_\mu g_\mu^* a_\mu$ and for $H'$ we have $c=ga_1$. If the initial state of the optical environment is taken as the vacuum, the only non-zero environmental correlation function is $\ev*{c(\tau)c^\dagger}$, where the expectation value $\ev{\cdot}$ is taken with respect to the free evolution of the environment. Following Ref.~\cite{si:tamascelli2018nonperturbative}, the equivalence between the emitter dynamics generated by $H$ and $H'$ is quantified as the similarity between the two correlation functions
\begin{align}
  \label{si:eq:1}
\begin{split}
  \Lambda(\tau) &= \ev{\sum_{\mu\mu'} g_\mu^*g_{\mu'} a_\mu(\tau)a_{\mu'}^\dagger} = \sum_\mu \abs{g_\mu}^2 \ev*{a_\mu(\tau)a_{\mu}^\dagger},
\\ \Lambda'(\tau) &= g^2\ev*{a_1(\tau)a_1^\dagger}.
\end{split}
\end{align}
Equivalently, their corresponding spectral densities, $J(\omega)=\int_{-\infty}^\infty\dd{\tau}\Lambda(\tau)e^{i\omega\tau}$ (and similarly for $J'$ and $\Lambda'$) can be used instead. The spectral density of the original system, $J$,  can be calculated as the LDOS from Sec.~\ref{sec:local-density-states}. For the mapped system, we can calculate $\Lambda'(\tau)$ using $H'$ and by tracing out the auxiliary reservoir. Imposing the Markov approximation on the auxiliary reservoir, the master equation corresponding to Eq.~\eqref{si:eq:99} is~\cite{si:breuer2002theory}
\begin{align}
  \label{si:eq:2}
\begin{split}
  \dot{\rho}(t)=-i&[H_\mathrm{E} + \sum_i\omega_ia_i^\dagger a_i+  (g a_1\sigma^\dagger + V a_1^\dagger a_2 + \mathrm{H.c.}),\rho(t)] \\ &+ \kappa\mathcal{D}[q_1a_1+q_2a_2],
\end{split}
\end{align}
where $\mathcal{D}(x)=x\rho(t) x^\dagger - \frac{1}{2}(x^\dagger x\rho(t)+\rho(t) x^\dagger x)$ and $\kappa(\omega)=2\pi\sum_\mu\abs{f_\mu}^2\delta(\omega-\epsilon_\mu)$ taken frequency-independent, consistent with the Markov approximation.

To calculate the environmental correlation function $\Lambda'(\tau)$, we simply set $H_\mathrm{E}=0, g=0$. Since the initial state of the environment is taken as the vacuum state, the system only explores a subspace of the two-cavity Fock space $\{\ket{n_1,n_2}\}$ spanned by the basis $\{\ket{0,0},\ket{1,0},\ket{0,1}\}$. In this basis, the Liouvillian corresponding to the RHS of the master equation, Eq.~\eqref{si:eq:2}, can be represented by a $9\times 9$ matrix, which we shall denote by $\mathcal{L}$. Since the interaction with the auxiliary environment is taken Markovian, we can calculate $\Lambda'(\tau)$ using the quantum regression theorem~\cite{si:breuer2002theory}, allowing for analytic calculation of the corresponding spectral density by matrix exponentiation,
\begin{widetext}
\begin{align}
  \label{si:eq:125}
  \begin{split}
    J'(\omega)&=2\abs{g}^2\Re{\int_0^\infty\dd{\tau}e^{i\omega\tau}\Tr[a_1 e^{\mathcal{L}\tau}(a_1^\dagger \dyad{0,0})]}\\
&=2\abs{g}^2\Re\qty[\frac{2i(\omega-\omega_2)-\kappa\abs{q_2}^2}{2(\omega-\omega_2)(\omega-\omega_1+\frac{1}{2}i\kappa\abs{q_1}^2)+(-2V+iq_2q_1^*\kappa)V^* + i\kappa[Vq_1q_2^*+\abs{q_2}^2(\omega-\omega_1)]}]
\\ &=2\abs{g}^2\Re\qty[\frac{2i(\omega-\omega_2)-\kappa_2}{2(\omega-\omega_2)(\omega-\omega_1)+i\kappa_1(\omega-\omega_2)+i\kappa_2(\omega-\omega_1)+2iV_0\sqrt{\kappa_1\kappa_2}\cos\varphi-2V_0^2}],
  \end{split}
\end{align}
\end{widetext}
where in the last equality, we have have defined 
\begin{align}
  \label{si:eq:126}
  \begin{split}
q_1&=:\abs{q_1}e^{i\phi_1}, \;
q_2=:\abs{q_2}e^{i\phi_2} \\
\kappa_1&:=\kappa\abs{q_1}^2,\;
\kappa_2:=\kappa\abs{q_2}^2,\\
V&=:V_0e^{i\theta},\;
\varphi:=\theta+\phi_1-\phi_2.
  \end{split}
\end{align}
In particular, we see that the spectral density does not depend on all of the complex phases individually, but only on the phase difference $\varphi$. We might thus conveniently set $\phi_1=\phi_2=0$ and thus $\theta=\varphi$. In this case, the dissipator in the master equation, Eq.~\eqref{si:eq:2}, simplifies as $\kappa\mathcal{D}[q_1a_1+q_2a_2]\rightarrow\mathcal{D}[\sqrt{\kappa_1}a_1+\sqrt{\kappa_2}a_2]$.

\subsubsection*{Analytic structure}
\label{sec:analytic-structure}

The two poles of $J'$, denoted by $z_\pm'$, are found as the roots of the denominator in \eqref{si:eq:125},
\begin{align}
  \label{si:eq:128}
  \begin{split}
    z_\pm'&=\frac{1}{2}\qty[2\Omega-\frac{i}{2}(\kappa_1+\kappa_2)]\pm\sqrt{D},
    \\
    D&=V_0^2-\qty[\frac{\kappa_1+\kappa_2}{4}]^2+\frac{i\Delta}{4}(\kappa_1-\kappa_2)+\frac{\Delta^2}{4} \\&-i\sqrt{\kappa_1\kappa_2}V_0\cos\varphi,
  \end{split}
\end{align}
where $\Omega:=\frac{1}{2}(\omega_1+\omega_2)$ and $\Delta:=\omega_2-\omega_1$.
This allows us to write $J'$ as 
\begin{align}
  \label{si:eq:129}
  J'(\omega)&=2\abs{g}^2 \Re\qty{\frac{i(\omega-\omega_2)-\frac{1}{2}\kappa_2}{(\omega-z_+')(\omega-z_-')}}.
\end{align}
From this form, we see that $z_\pm'$ are simple poles, and the corresponding residues, $R_\pm'$, are thus
\begin{align}
  \label{si:eq:130}
  \begin{split}
    R_\pm'&=\lim_{\omega\rightarrow z_\pm'} (\omega-z_\pm')\mathcal{J}(\omega), \\
    &= \abs{g}^2\frac{i(z_\pm'-\omega_2)-\frac{1}{2}\kappa_2}{z_\pm'-z_\mp'}
  \end{split}
\end{align}

\subsection{Determining the parameters of the mapped environment}
\label{sec:determ-param-mapp}
\begin{figure}
  \centering
  \includegraphics[width=\columnwidth]{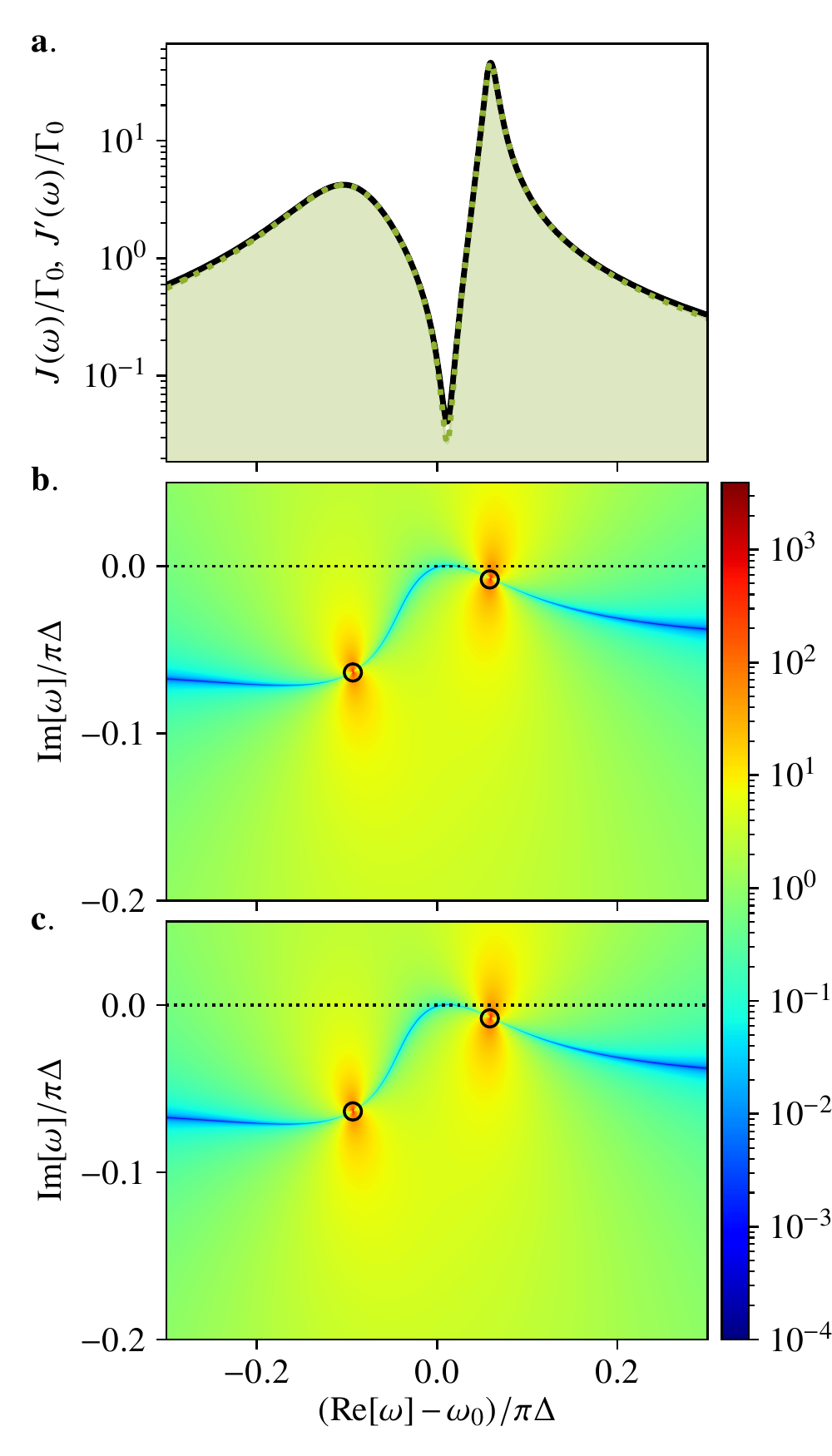}
  \caption{{\bf a.} Original (black solid) and mapped (green dotted, shaded area) spectral densities, $J$ and $J'$ as functions of real frequency, normalised to the bare spontaneous emission rate, $\Gamma_0$. {\bf b-c.} Analytic continuation of spectral densities, $\abs{J}/\Gamma_0$ (b) and $\abs{J'}/\Gamma_0$ (c). The black circles indicate the poles. Parameters: $\gamma_0=10^-3\Delta,\; r_\mathrm{B}=-1/\sqrt{2},\; \omega_0=3\Delta,\; \omega_\mathrm{F}=\omega_0-0.02\pi\Delta,\; P=1,\; \gamma_1=\gamma_2=0.05\Delta.$}
  \label{si:fig:analytic-structure}
\end{figure}

Rather than optimising all of the seven parameters $(g,\kappa_1,\kappa_2,V_0,\theta,\omega_1,\omega_2)$ freely using a numerical fitting routine, we exploit our access to analytical properties of $J$ and $J'$.

By calculating the analytical continuation of $J$ in the complex plane in the vicinity of the Fano cavity resonance, two poles can be found (see Fig.~\ref{si:fig:analytic-structure}), which we shall denote by $z_\pm$. By numerically integrating $J(z)$ along a closed contour, $C_\pm$, enclosing a single pole, $z_\pm$, the residue of this pole, $R_\pm$, can be calculated using the residue theorem as 
\begin{align}
  \label{si:eq:3}
  R_\pm=\frac{1}{2\pi i}\oint_{C_\pm} \dd{z} J(z).
\end{align}
Choosing the contour $C_\pm$ to be a circle with radius $Z_0$, $C_\pm=\{z:\phi\rightarrow Z_0e^{i\phi}+z_\pm\:|\:\phi\in[0,2\pi]\}$, the residue becomes
\begin{align}
  \label{si:eq:4}
  \frac{Z_0}{2\pi}\int_0^{2\pi}\dd{\phi}J(Z_0e^{i\phi}+z_\pm).
\end{align}

We now require 
\begin{align}
  \label{si:eq:6}
z_+'=z_+,\; z_-'=z_-.
\end{align}
Adding these two equations yields
\begin{align}
  \label{si:eq:7}
  \Omega&=\frac{1}{2}\Re[z_++z_-],\\
\label{si:eq:8}
\kappa_1+\kappa_2&=-2\Im[z_++z_-].
\end{align}
Subtracting the two equations \eqref{si:eq:6} yields $2\sqrt{D}=z_+-z_-$. Squaring and taking the real and imaginary parts of this equation leads to
\begin{align}
  \label{si:eq:9}
  &V_0^2-\qty(\frac{\kappa_1+\kappa_2}{4})^2+\frac{\Delta^2}{4}=\frac{1}{4}\Re[(z_+-z_-)^2], \\
\label{si:eq:10}
&\frac{\Delta}{4}(\kappa_1-\kappa_2)-\sqrt{\kappa_1\kappa_2}V_0\cos\varphi=\frac{1}{4}\Im[(z_+-z_-)^2].
\end{align}
Proceeding similarly with the residues $R_\pm$ and $R_\pm'$ would lead to four additional constraints, which would overdetermine the system. Alternatively, we use the restriction $\Im[R_+'+R_-']=\Im[R_++R_-]$ to generate the constraint
\begin{align}
  \label{si:eq:11}
  g=\sqrt{\Im[R_++R_-]}.
\end{align}
The equations \eqref{si:eq:7}-\eqref{si:eq:11} gives five constraints, leaving two degrees of freedom. In our numerical implementation, we leave the two parameters $\Delta$ and $\kappa_2$ as free. We then choose the pair $(\Delta,\kappa_2)$ that minimises error function $\epsilon=\int_W \dd{\omega} [J(\omega)-J'(\omega)]^2$, where the integral runs over a frequency window, $W$ that covers the central features of the LDOS in the vicinity of the resonance.

\section{Polaron master equation and dynamics}
\label{sec:polar-mast-equat}

\subsection{Master equation}
\label{sec:master-equation}

The full Hamiltonian of the mapped optical environment and phonon reservoir is $H'+H_\mathrm{P}+H_\mathrm{EP}$, where $H'$ is as in Eq.~\eqref{si:eq:99}, $H_\mathrm{E}=\omega_{eg}\sigma^\dagger\sigma$ describes the free evolution of the emitter, $H_\mathrm{P}=\sum_\mathbf{q}\nu_\mathbf{q}b_\mathbf{q}^\dagger b_\mathbf{q}$ describes the free evolution of the phonons and $H_\mathrm{EP}=\sigma^\dagger\sigma\sum_\mathbf{q}M_\mathbf{q}(b_\mathbf{q}+b_\mathbf{q}^\dagger)$ describes the emitter--phonon coupling. In the polaron frame, described by the unitary transformation $U_\mathrm{P}=\dyad{g}+\dyad{e}B_+$, with $B_\pm=\exp[\pm\sum_\mathbf{q}\nu_\mathbf{q}^{-1}M_\mathbf{q}(b_\mathbf{q}^\dagger-b_\mathbf{q})]$, the Hamiltonian is 
$\hat{H}':=U_\mathrm{P}H'U_\mathrm{P}^\dagger=:\hat{H}_0+\hat{H}_\mathrm{R}+\hat{H}_\mathrm{I}$, where 
\begin{align}
  \label{si:eq:11}
  \hat{H}_0&= \hat{\omega}_{eg}\sigma^\dagger\sigma + \sum_i \omega_ia_i^\dagger a_i + \hat{g}X+(Va_1^\dagger a_2 + V^*a_2^\dagger a_1)
\end{align}
describes the internal dynamics of the two-level system and the two mapped optical modes,
\begin{align}
  \label{si:eq:13}
  \hat{H}_\mathrm{R}=\sum_\mu \epsilon_\mu \alpha_\mu^\dagger \alpha_\mu + \sum_\mathbf{q} \nu_\mathbf{q} b_\mathbf{q}^\dagger b_\mathbf{q}
\end{align}
describes the free evolution of the total reservoir comprising the auxiliary electromagnetic environment and the phonon environment, and
\begin{align}
  \label{si:eq:14}
\begin{split}
  \hat{H}_\mathrm{I}&=[q_1a_1^\dagger + q_2 a_2^\dagger]\sum_\mu f_\mu \alpha_\mu + [q_1^*a_1 + q_2^* a_2]\sum_\mu f_\mu^* \alpha_\mu^\dagger \\
&+g(XB_X+YB_Y)
\end{split}
\end{align}
describes the system-reservoir interactions. The quantities entering these expressions are defined as $X=\sigma^\dagger a_1+\sigma a_1^\dagger,\; Y=i(\sigma^\dagger a_1-\sigma a_1^\dagger),\; B_X=(B_++B_--2B_0)/2,\; B_Y=i(B_+-B_-)/2,\; \hat{g}=B_0g$, where $B_0=\Tr[\rho_\mathrm{P}^0B_+]$ is the expectation value of $B_+$ with respect to the phonon thermal state, $\rho_\mathrm{P}^0=\exp[-\frac{1}{k_\mathrm{B}T}\sum_\mathbf{q}\nu_\mathbf{q} b_\mathbf{q}^\dagger b_\mathbf{q}]/\Tr{\exp[-\frac{1}{k_\mathrm{B}T}\sum_\mathbf{q}\nu_\mathbf{q} b_\mathbf{q}^\dagger b_\mathbf{q}]}$, with $k_\mathrm{B}$ and $T$ the Boltzmann constant and temperature, respectively. Tracing out the phonon environment and auxiliary electromagnetic environment leads to the second order Born-Markov master equation for the reduced density operator, $\rho$~\cite{si:breuer2002theory}
\begin{align}
  \label{si:eq:17} 
\begin{split}
\dot{\rho}(t)=&-i[\hat{H}_0,\rho(t)]\\
&-\int_0^\infty\dd{\tau}\Tr_\mathrm{R}[H_\mathrm{I},[H_\mathrm{I}(-\tau),\rho(t)\otimes\mathcal{\rho}_\mathrm{P}^0\otimes\rho_\xi^0]],
\end{split}
\end{align}
where the reference state of the auxiliary electromagnetic environment is taken as the vacuum, $\rho_\xi^0=\bigotimes_\mu \dyad{0_\mu}$, assuming temperatures significantly below the typical optical frequencies around the emitter resonance, $k_\mathrm{B}T\ll \omega_{eg}$. Writing this master equation out explicitly yields
\begin{align}
  \label{si:eq:15}
  \dot{\rho}(t)=-i[\hat{H}_0,\rho(t)]+\mathcal{D}[\sqrt{\kappa_1}a_1+\sqrt{\kappa_2}a_2]+\mathcal{W},
\end{align}
where the phonon-induced term is 
\begin{align}
  \label{si:eq:16}
\mathcal{W}=g^2([X,\rho(t) \Theta_X^\dagger]+[Y,\rho(t)\Theta_Y^\dagger]+\mathrm{H.c.}),
\end{align}
where $\Theta_\zeta=\int_0^\infty\dd{t}\zeta(-\tau)\Lambda_\zeta(\tau)$, with $\zeta=X,Y$ and $\zeta(-\tau)$ denoting the free Heisenberg picture time evolution, $\zeta(-\tau)=e^{-i\hat{H}_0\tau}\zeta e^{+i\hat{H}_0\tau}$. The phonon correlation functions are $\Lambda_X(\tau)=\frac{1}{2}B^2[e^{\phi(t)}+e^{-\phi(\tau)}-2],\; \Lambda_Y(\tau)=\frac{1}{2}B^2[e^{\phi(\tau)}-e^{-\phi(\tau)}]$ with $\phi(\tau)=\int_0^\infty\dd{\nu}\nu^{-2}J_\mathrm{P}(\nu)[\coth(\beta\nu/2)\cos\nu\tau-i\sin\nu\tau]$ and $\beta=1/(k_\mathrm{B}T)$. The phonon spectral density is $J_\mathrm{P}(\nu)=\sum_\mathbf{q}M_\mathbf{q}^2\delta(\nu-\nu_\mathbf{q})=\alpha \nu^3 \exp[-\nu^2/\nu_c^2]$, with overall coupling strength $\alpha$ and cutoff frequency $\nu_c$.

\subsection{Dipole spectrum}
\label{sec:dipole-spectrum-from}
When calculating dynamics in the polaron frame, quantities of interest must be transformed back to the lab frame. In particular, the dipole correlation function, $\ev{\sigma^\dagger(t)\sigma(t')}_\mathrm{LF}$ (with subscript $\mathrm{LF}$ signifying expectation values in the lab frame) is calculated from the polaron frame (subscript $\mathrm{PF}$) dynamics as~\cite{si:roy2012polaron}
\begin{align}
  \label{si:eq:18}
  \ev{\sigma^\dagger(t)\sigma(t')}_\mathrm{LF}=\ev{\sigma^\dagger(t)B_+(t)B_-(t')\sigma(t')}_\mathrm{PF}.
\end{align}
Assuming that the two-level system and phonons are weakly coupled in the polaron frame, this is approximated as 
\begin{align}
  \label{si:eq:19}
  \ev{\sigma^\dagger(t)\sigma(t')}_\mathrm{LF} \simeq  \ev{B_+(t)B_-(t')}_\mathrm{PF}\ev{\sigma^\dagger(t)\sigma(t')}_\mathrm{PF},
\end{align}
and $\ev{B_+(t)B_-(t')}_\mathrm{PF}$ is approximated as the equilibrium phonon correlation function, $\ev{B_+(t)B_-(t')}_\mathrm{PF}\simeq B_0^2e^{\phi(t-t')}$. 

The two-colour emission spectrum, $S(\omega,\omega')$, observed in the waveguide region on the right hand side of the Fano mirror is related to the dipole correlation function as~\cite{si:roy2015quantum,si:iles2017phonon}
\begin{align}
  \label{si:eq:21}
  S(\omega,\omega')=G(\omega)G(\omega')S_0(\omega,\omega'),
\end{align}
where $S_0(\omega,\omega')=\int_{-\infty}^\infty\dd{t}\dd{t'}\ev{\sigma^\dagger(t)\sigma(t')}_\mathrm{LF}e^{i(\omega t-\omega't')}$ is the dipole spectrum and 
\begin{align}
  \label{si:eq:22}
  G(\omega)=\frac{(1+r_0e^{i\omega/\Delta})t_\mathrm{F}(\omega)}{1-r_0r_\mathrm{F}(\omega)e^{2i\omega/\Delta}}
\end{align}
is the Green's function that connects the dipole spectrum with the spectrum of the electromagnetic field in the waveguide, transmitted through the Fano mirror~\cite{si:denning2018cavity}.

\section{Equivalent Fabry-Perot cavity}
\label{sec:equiv-fabry-perot}

\begin{figure}
  \centering
  \includegraphics[width=\columnwidth]{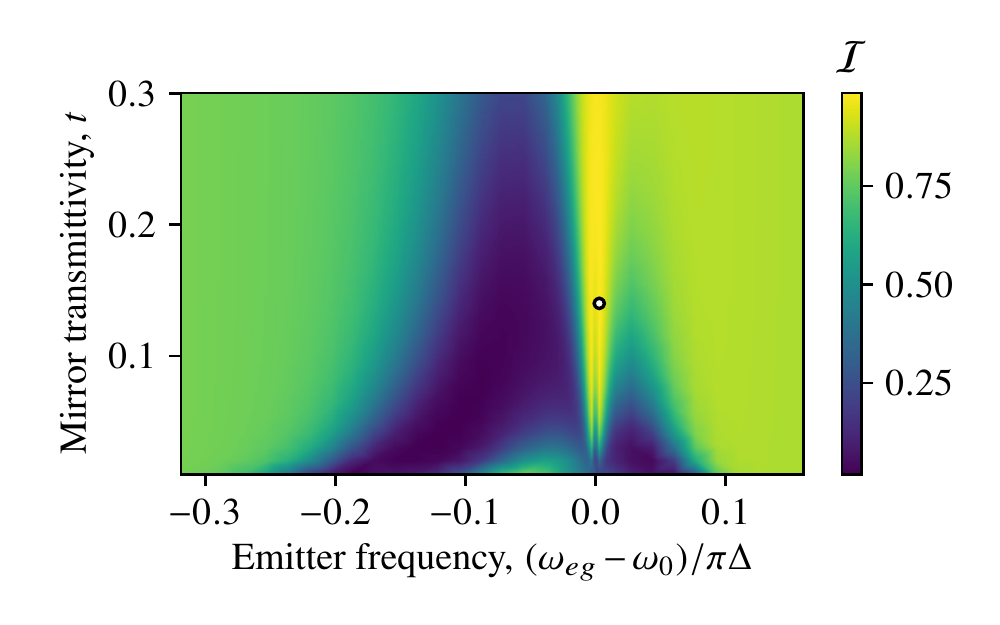}
  \caption{Indistinguishability of Fabry-Perot cavity as a function of emitter frequency and right mirror transmittivity. Parameters: $\Delta=10\mathrm{\;meV},\; \Gamma_0=0.6\mathrm{\; \mu eV},\; \Gamma_\mathrm{R}=0.03\mathrm{\: \mu eV},\; \alpha=0.03\mathrm{\; ps^2},\; \nu_c=1.45\mathrm{\: meV},\; T=4\mathrm{\: K}.$ The white dot indicates the minimal distinguishability, $\delta=0.013$.}
  \label{si:fig:FP-indist}
\end{figure}

When both mirrors in the cavity are regular broadband mirrors, a simpler analysis can be carried out. The LDOS in the middle position of the cavity is then
\begin{align}
  \label{si:eq:12}
  J(\omega)=\Gamma_0 \Re\qty[\frac{(1+r_0e^{i\omega/\Delta)})(1+re^{i\omega/\Delta})}{1-r_0re^{2i\omega/\Delta}}],
\end{align}
where $r$ is the reflectivity of the right mirror. Following the strategy from Ref.~\cite{si:denning2018cavity}, a single cavity mode can be extracted from the LDOS with reflectivity-dependent emitter coupling strength, $g$, and decay rate, $\kappa$. The master equation describing the dynamics (including interactions with phonons) is then
\begin{align}
  \label{si:eq:20}
  \dot{\rho}(t)=-i[B_0g(a\sigma^\dagger+a^\dagger\sigma),\rho(t)]+\kappa L(a)+\mathcal{W},
\end{align}
where $a$ is the annihilation operator of the cavity mode and $\mathcal{W}$ is as in Eq.~\eqref{si:eq:16}, but with $X=a\sigma^\dagger+a^\dagger\sigma,\; Y=i(a\sigma^\dagger-a^\dagger\sigma)$. The emission spectrum and indistinguishability is calculated using the same technique as for the Fano cavity, but with the Fabry-P\'erot Green's function
\begin{align}
  \label{si:eq:24}
  G(\omega)=\frac{(1+r_2e^{i\omega/\Delta})t}{1-r_0re^{2i\omega/\Delta}},
\end{align}
where $t=\sqrt{1-r^2}$ is the transmittivity of the right mirror.

Fig.~\ref{si:fig:FP-indist} shows the indistinguishability for a Fabry-P\'erot cavity as a function of emitter frequency and transmittivity of the right mirror. Besides from the right mirror, all the properties of the optical structure are the same as in Fig. 3 of the main text. The minimal distinguishability of 0.013 is indicated with a dot.

\end{document}